\documentclass[preprint,authoryear,12pt]{elsarticle}
\usepackage[top=3cm, bottom=3cm, left=3cm, right=3cm]{geometry}
\usepackage{amsbsy,amssymb,amsmath,bm}
\usepackage[font=small,labelfont=bf]{caption}
\usepackage{graphicx,subfig}
\usepackage[authoryear]{natbib}
\usepackage{comment}
\usepackage[draft]{changes}

\makeatletter
\providecommand\phantomcaption{\caption@refstepcounter\@captype}
\makeatother

\journal{Physics of the Earth and Planetary Interiors}

\begin{document}

\begin{frontmatter}

\title{A simplified model of collision-driven dynamo action in small bodies}
\author{Xing Wei$^1$, Rainer Arlt$^2$, Andreas Tilgner$^1$}
\address{$^1$Institute of Geophysics, University of G\"ottingen, Friedrich-Hund-Platz 1, 37077 G\"ottingen, Germany\\
$^2$Leibniz Institute of Astrophysics Potsdam, An der Sternwarte 16, 14482 Potsdam, Germany}
\date{\today}

\begin{abstract} 
We investigate numerically the self-sustained dynamo action in a spinning sphere whose sense of rotation reverses periodically. This system serves as a simple model of a dynamo in small bodies powered by frequent collisions. It is found that dynamo action is possible in some intervals of collision rates. At high Ekman numbers the laminar spin-up flow is helical in the boundary layers and the Ekman circulation together with the azimuthal shear powers the dynamo action. At low Ekman number a non-axisymmetric instability helps the dynamo action. The intermittency of magnetic field occurs at low Ekman number.
\end{abstract}

\begin{keyword}
dynamo, magnetohydrodynamics, spin-up, planetesimals
\end{keyword}

\end{frontmatter}

\section{Motivation} 
Magnetic fields are ubiquitous in the universe and dynamo action is believed to create those magnetic fields by transforming kinetic energy of a fluid electric conductor into magnetic energy. The fluid motion is probably driven by thermal and/or compositional buoyancy in the Earth's fluid core and in the solar convection zone. Mechanical forcing is an alternative driving mechanism. \citet{bullard} once pointed out that the Earth's precession is capable to drive the geodynamo and \citet{andreas} implemented numerically a precession dynamo. Recently, several mechanically forced dynamos were proposed, such as the Martian dynamo induced by tidal forces \citep{aldridge}, the lunar dynamo induced by precession \citep{dwyer} and collisions \citep{bars}, and dynamos in planetesimals induced by collisions \citep{weiss}. However, these recent works focused on the planetary science and did not verify by solving the equations of magnetohydrodynamics (MHD) that the forcing under consideration leads to a flow capable of dynamo action.

An irregularly shaped small body, e.g. planetesimal with a liquid core will in general execute a complicated motion following a collision, with precession and nutation of its rotation axis. In this paper, we simplify this problem geometrically by considering a spherical core, and dynamically by assuming that the planetesimal has a constant angular velocity between collisions and that the effect of collision is to reverse the direction of rotation. The fluid inside the sphere thus undergoes a sequence of spin-up motions whose property as
a dynamo we investigate. In section 2 we formulate the equations and introduce the numerical method. In section 3 we present our computational results. In section 4 we summarise our results and propose an application to planetary science.

\section{Equations}
We consider an incompressible conducting fluid in a spherical shell with the outer sphere rotating about the symmetric axis back and forth due to tangential collisions. The dimensionless Navier-Stokes equation in the inertial frame reads
\begin{equation}
\label{ns} 
\frac{\partial\bm U}{\partial t}+\bm U\cdot\bm\nabla\bm U=-\bm\nabla P+Re^{-1}\nabla^2\bm U+(\bm\nabla\bm\times\bm B)\times\bm B, 
\end{equation} 
and the dimensionless magnetic induction equation reads 
\begin{equation}
\label{induction} 
\frac{\partial\bm B}{\partial t}=\bm\nabla\times(\bm U\times\bm B)+Rm^{-1}\nabla^2\bm B.  
\end{equation}
In the governing equations (\ref{ns}) and (\ref{induction}) length is normalised with the outer radius $r_o$, time with the rotation time scale of the outer sphere $\Omega^{-1}$, velocity with $\Omega r_o$, and magnetic field with $\sqrt{\rho\mu}\Omega r_o$, where $\rho$ is the fluid density and $\mu$ the magnetic permeability. There are then two dimensionless parameters, the Reynolds number $Re$ and the magnetic Reynolds number $Rm$. The Reynolds number 
\begin{equation}
Re=\frac{\Omega r_o^2}{\nu} 
\end{equation} 
measures the rotation of the outer sphere which is the driving agent, and is the inverse of the Ekman number
\begin{equation} 
Ek=\frac{\nu}{\Omega r_o^2}, 
\end{equation} 
which is the ratio of the rotational time scale to the viscous time scale. The magnetic Reynolds number
\begin{equation} 
Rm=\frac{\Omega r_o^2}{\eta} 
\end{equation} 
measures the ratio of the magnetic induction to the magnetic diffusion, which is equivalent to the ratio of the magnetic diffusive time scale to the fluid advective time scale. In our computations the aspect ratio is fixed to be $r_i/r_o=0.1$ such that the inner sphere is small enough to be negligible. The velocity boundary condition is no-slip at $r=r_o$ and stress-free at $r=r_i$ to approximate a full sphere. The boundary condition at $r_o$ for the reversals of rotation direction is as follows: 
\begin{align} 
&{\rm if ~~ mod}(IP(t/\tau),2)=0 ~~
{\rm then} ~~ u_\phi=\sin\theta \hspace{3mm} &\text{at} \hspace{3mm} r=r_o, \nonumber\\ 
&{\rm if ~~ mod}(IP(t/\tau),2)=1 ~~ {\rm then} ~~ u_\phi=-\sin\theta \hspace{3mm} &\text{at} \hspace{3mm} r=r_o.  
\end{align}
where $IP(x)$ denotes the integer part of a real number $x$, $\tau$ is the length of the interval between collisions and $\theta$ is the colatitude. The magnetic boundary condition is insulating at $r_o$ and $r_i$. The initial field is a random seed field. This boundary condition seems like a longitudinal libration and the librating flows have been extensively studied, e.g. \citet{calkins}, \citet{zhang}, and \citet{sauret}. Libration is a small harmonic motion superposed on a global rotation. Our boundary condition is the Heaviside function in time in respect of rotation direction.

The numerical method is a standard pseudo-spectral method in spherical coordinates $(r,\theta,\phi)$ \citep{rainer}. The nonlinear terms are evaluated in physical space and transformed back and forth between physical and spectral spaces. The toroidal-poloidal decomposition method is employed for fluid flow and magnetic field such that the divergence-free condition $\bm\nabla\bm\cdot\bm u=\bm\nabla\bm\cdot\bm B=0$ is automatically satisfied. The fluid flow and magnetic field are decomposed as
\begin{align} 
\bm U=\bm\nabla\times (e\hat{\bm r})+\bm\nabla\times\bm\nabla\times (f\hat{\bm r}), \nonumber\\ 
\bm B=\bm\nabla\times (g\hat{\bm r})+\bm\nabla\times\bm\nabla\times (h\hat{\bm r}), \label{t-p}
\end{align} 
where $e$ and $f$ are respectively the toroidal and poloidal components of $\bm U$, $g$ and $h$ are respectively the toroidal and poloidal components of $\bm B$, and $\hat{\bm r}$ denotes the unit vector in the radial direction. The spherical harmonics $P_l^m(\cos\theta)e^{{\mathrm i}m\phi}$ are used on the spherical surface $(\theta,\phi)$ and the Chebyshev polynomials $T_k(r)$ are used in the radial direction. For example, $e$ is represented by
\begin{equation} 
e=\sum_{k,l,m}e_{klm}(t)T_k(r)P_l^m(\cos\theta)e^{{\mathrm i}m\phi}, 
\end{equation} 
where $k$, $l$ and $m$ are respectively the radial, colatitude and longitude wavenumbers. The second order Runge-Kutta method is used for the time stepping. The resolutions up to $k=128$ Chebyshev modes in the radial direction, $l=128$ Legendre modes in the colatitude direction and $m=32$ Fourier modes in the longitude direction are used.

\section{Results} 
The evolution of an unsteady flow driven by spinning boundaries is called the spin-up process \citep{greenspan}. In the spin-up process the fluid motion develops near the boundary forming Ekman layers, which then causes a secondary flow, the Ekman pumps, emerging from the boundary layers and extending through the whole fluid interior. The spin-up process ultimately leads to a solid body rotation. The time scale of the spin-up process, the thickness of the boundary layer and the velocity of Ekman pumps are respectively proportional to $Ek^{-1/2}$, $Ek^{1/2}$ and $Ek^{1/2}$. The solid body rotation in a sphere is a toroidal flow and cannot act as a dynamo. If, however, the spin-up is excited anew at regular intervals by a modification of the rotation of the boundaries, we can expect dynamo action arising from this time-dependent flow.

\begin{figure}
\centering
\caption{Dynamo windows at (a) $Re=2\times10^{2}$ or $Ek=5\times10^{-3}$, (b) $Re=5\times10^{2}$ or $Ek=2\times10^{-3}$, (c) $Re=1\times10^{3}$ or $Ek=1\times10^{-3}$ and (d) $Re=2\times10^{3}$ or $Ek=5\times10^{-4}$. The horizontal axis shows $Rm$ and the vertical axis shows the collision interval. Squares denote dynamos and crosses failed dynamos.}\label{rev}
\begin{minipage}{0.5\textwidth}
\centering
\includegraphics[scale=0.45]{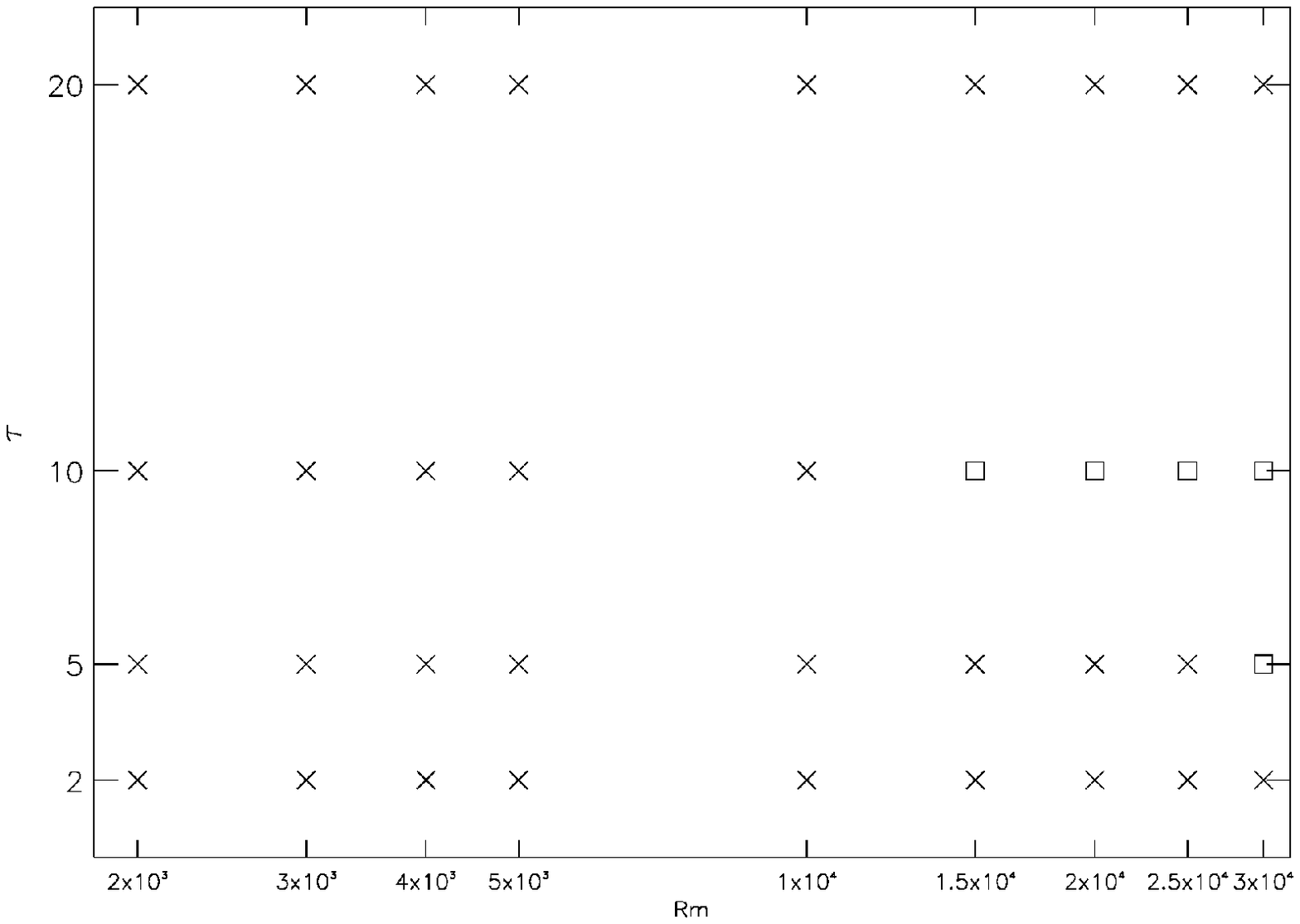}\label{rev1}
\caption*{(a) $Re=2\times10^{2}$ or $Ek=5\times10^{-3}$}
\end{minipage}%
\begin{minipage}{0.5\textwidth}
\centering
\includegraphics[scale=0.45]{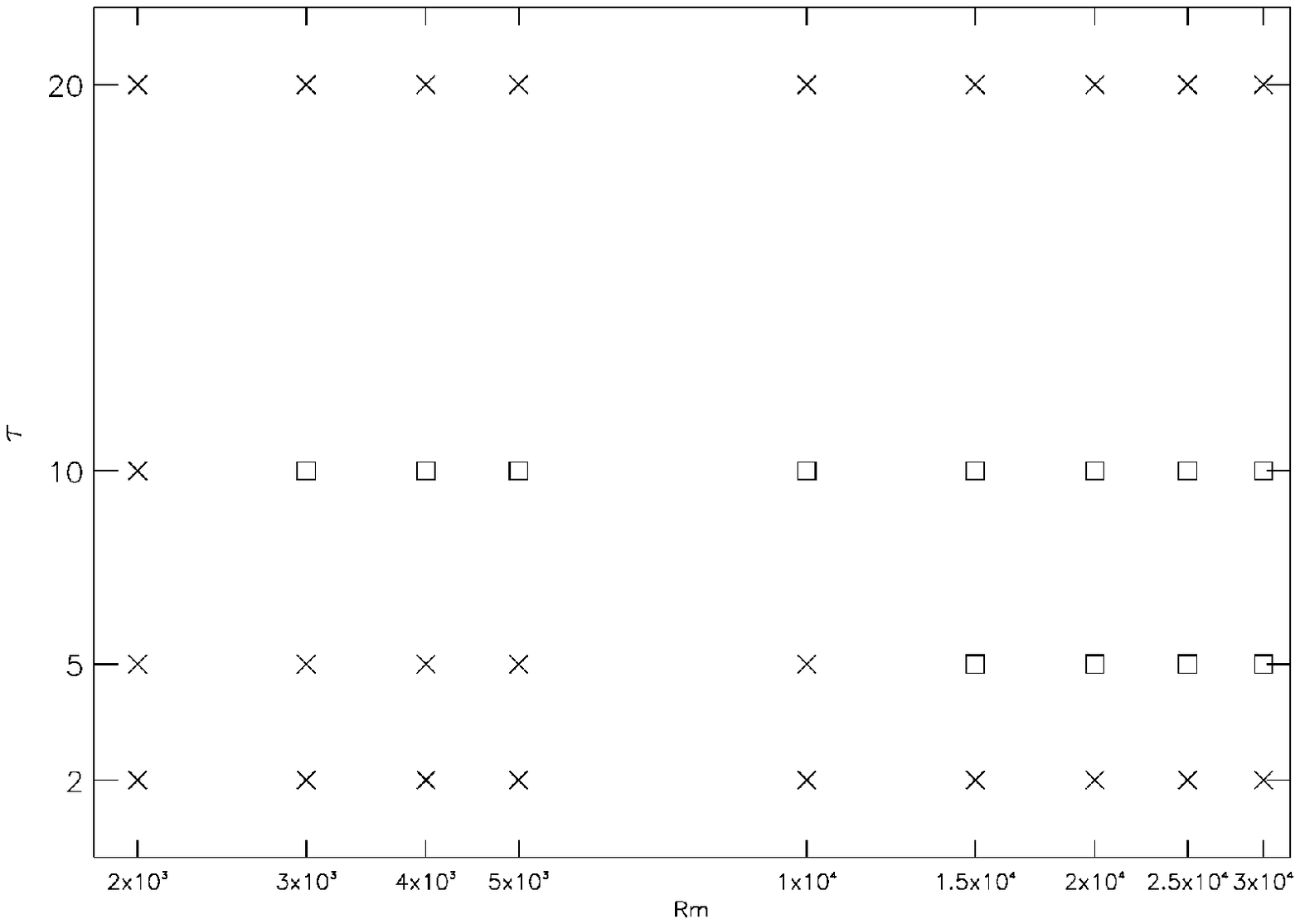}\label{rev2}
\caption*{(b) $Re=5\times10^{2}$ or $Ek=2\times10^{-3}$}
\end{minipage}
\begin{minipage}{0.5\textwidth}
\centering
\includegraphics[scale=0.45]{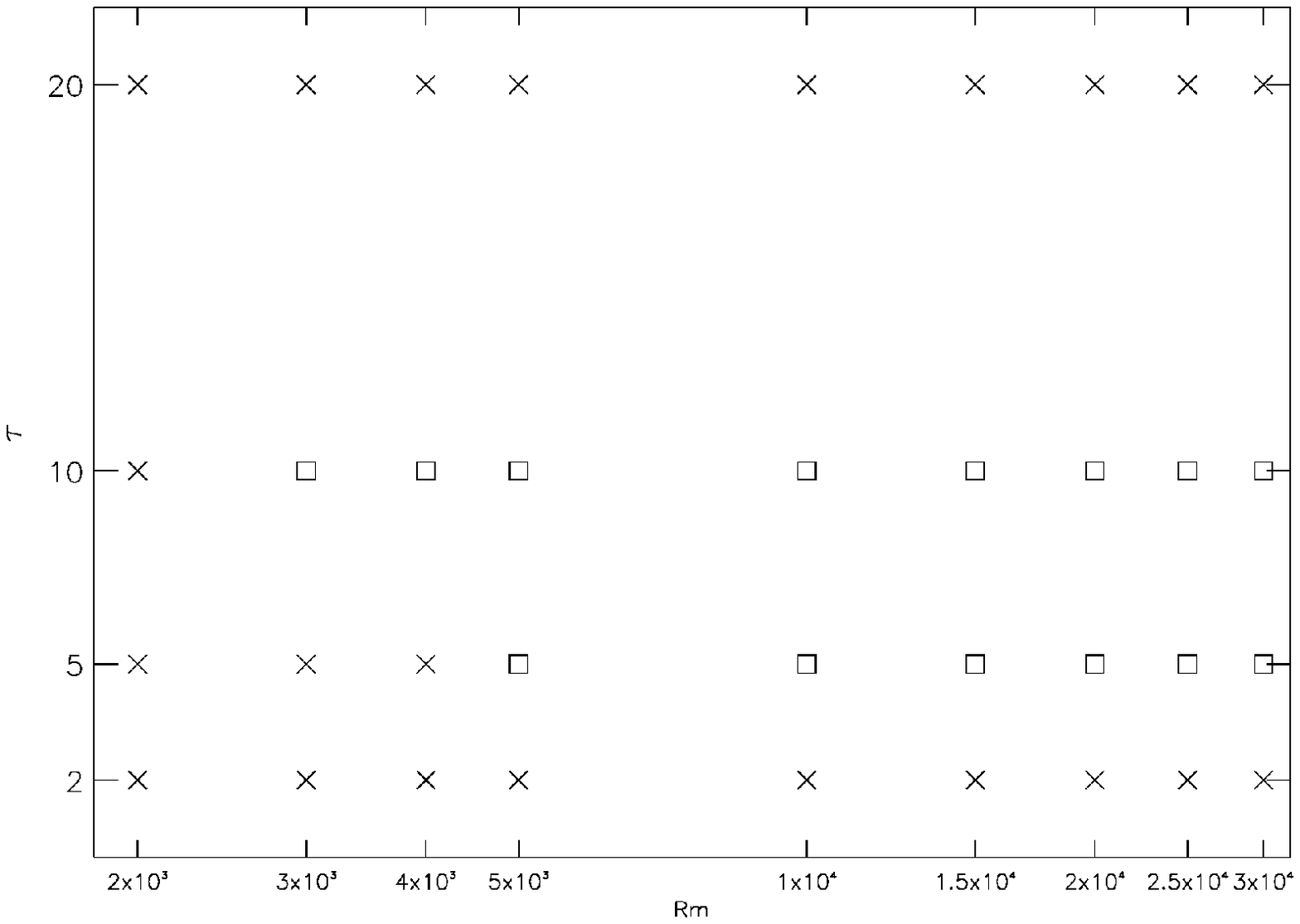}\label{rev3}
\caption*{(c) $Re=1\times10^{3}$ or $Ek=1\times10^{-3}$}
\end{minipage}%
\begin{minipage}{0.5\textwidth}
\centering
\includegraphics[scale=0.45]{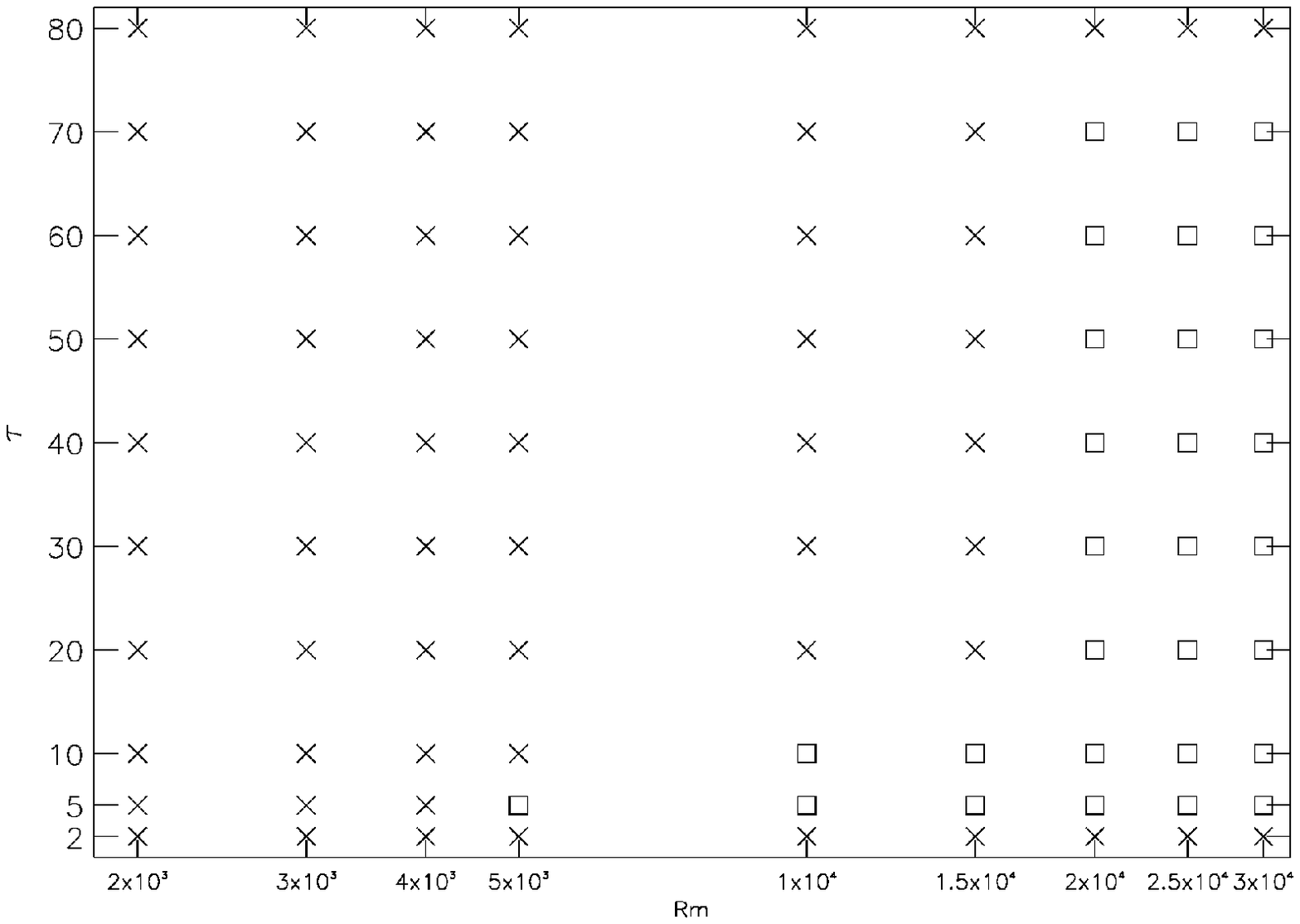}\label{rev4}
\caption*{(d) $Re=2\times10^{3}$ or $Ek=5\times10^{-4}$}
\end{minipage}
\end{figure}

We numerically simulate different combinations of $Re$ or $Ek$, $Rm$ and $\tau$ to search for dynamos.
Figure \ref{rev} shows successful and failed dynamos in the plane spanned by $Rm$ and $\tau$ at four different $Re$ or $Ek$, namely $Re=2\times10^{2}$, $5\times10^{2}$, $1\times10^{3}$, and $2\times10^{3}$ (or alternatively $Ek=5\times10^{-3}$, $2\times10^{-3}$, $1\times10^{-3}$, and $5\times10^{-4}$). In each figure at a fixed $Re$, a variation of $Rm$ is equivalent to a variation of magnetic Prandtl number, which is the ratio of viscosity to magnetic diffusivity. As expected, at fixed $Ek$ and $Rm$, dynamos occur only in a certain window or interval of $\tau$. If $\tau$ is too small and rotation reverses too often, the fluid in the interior of the sphere barely follows the motion of the boundaries and acquires too little kinetic energy. If on the other hand $\tau$ is too large and collisions are infrequent, the fluid spends much time in a state close to solid body rotation, which is not favorable for dynamo action. Dynamo action of the overall flow can thus only be expected for intermediate values of $\tau$. At the largest three $Ek$ investigated in figure \ref{rev}, dynamos are found at $\tau=5$ and $10$, but at the lowest $Ek$, the dynamos occur at $\tau$ up to $70$.

In the dimensionless equations we use $\Omega^{-1}$ to normalise time such that the dimensionless spin-up time scale is $Ek^{-1/2}$. If, instead, we use the dimensional spin-up time scale $Ek^{-1/2}\Omega^{-1}$ to normalise time, then the dimensionless $\tau'$ should be related to $\tau$ through $\tau'=Ek^{1/2}\tau$. $Ek$ in our computations is taken to be $5\times10^{-3}$, $2\times10^{-3}$, $1\times10^{-3}$ and $5\times10^{-4}$. Accordingly, $Ek^{1/2}$ is $0.071$, $0.045$, $0.032$ and $0.022$. If we translate $\tau$ to $\tau'$, then at the three higher $Ek$ the dynamos occur at $\tau'=0.016$, $0.225$, $0.032$, $0.355$, $0.045$, $0.710$ (sorted from the smallest to the largest), and at the lowest $Ek$ the dynamos occur at $\tau'$ ranging from $0.110$ to $1.540$. The maximum $\tau'$ at the lowest $Ek$ is almost twice of that at the three higher Ek. So the physics of a much wider dynamo window at the lowest $Ek$ is dynamically different from the physics of narrow windows at the three higher $Ek$. This spin-up time scale is usually applied to a laminar flow, but the flow at the lowest $Ek$ is already unstable (though not very turbulent, we will see later), and therefore the narrow dynamo windows at high $Ek$ results from a laminar flow whereas the much wider dynamo window at the lowest $Ek$ results from the flow instability.

Similarly, we can choose the Ekman pumping velocity which is at the order of $Ek^{1/2}\Omega r_o$ to normalise the velocity instead of $\Omega r_o$ to investigate the critical magnetic Reynolds number. $Rm_c$ at the four $Ek$ is $15000$, $3000$, $3000$ and $5000$. Then $Rm'_c$ is $Rm'_c=Ek^{1/2}Rm_c=1065$, $135$, $96$, $110$. $Rm'_c$ at the highest $Ek$ (or the lowest $Re$), corresponding to the strongest Ekman pumping, is highest, and this indicates that the Ekman pumping alone cannot be responsible for the dynamo action. On the other hand, at the highest $Ek$ (or the lowest $Re$), the azimuthal shear is the weakest and so $Rm'_c$ is the highest. Therefore, in a laminar flow at high $Ek$ both the Ekman pumping together with the azimuthal shear powers the dynamo action.

To better understand this spin-up dynamo, we need to know more about the structures of fluid flow and magnetic field. Figure \ref{mer} shows the axisymmetric fluid flow and magnetic field in the meridional plane. Subfigures \ref{mer1} and \ref{mer2} are at the higher $Ek=2\times10^{-3}$, and subfigures \ref{mer3} and \ref{mer4} at the lowest $Ek=5\times10^{-4}$. We first investigate the flow in subfigures \ref{mer1} and \ref{mer2}. The angular velocity of the fluid, given by $u_\phi/(r \sin \theta)$, exhibits in the interior the two-dimensional columnar structure demanded by the Taylor-Proudman theorem with a fluid motion independent of the coordinate along the rotation axis. The Ekman layer develops in the vicinity of outer boundary but not at the inner boundary because the stress-free condition is imposed at the inner boundary, and the meridional circulation concentrates near the outer boundary. Although the angular velocity reverses during the two successive collisions, the meridional circulation is always clockwise in the northern hemisphere and anti-clockwise in the southern hemisphere because the Ekman pumping due to collisions always drives the flow away from the equator along the axis of the sphere. We next investigate the magnetic field in subfigures \ref{mer1} and \ref{mer2}. The toroidal field is anti-symmetric about the equator, and it is concentrated near the outer boundary where the shear is strongest, and the poloidal field is dipolar. It is interesting that the dipolar field always points to the same direction while the toroidal field reverses after collisions. We then move to subfigures \ref{mer3} and \ref{mer4} to investigate the lowest $Ek$. Both the differential rotation and the meridional circulation are more concentrated in the thinner Ekman layer. The toroidal field becomes equatorially symmetric and the poloidal field is quadrupolar. This symmetry is also admitted by the magnetic induction equation, namely either an equatorially anti-symmetric toroidal field with a dipolar field (the two right panels in subfigures \ref{mer1} and \ref{mer2}) or an equatorially symmetric toroidal field with a quadrupolar field (the two right panels in subfigures \ref{mer3} and \ref{mer4}) can be the dynamo solution.

The spin-up flow has the same topology as the S2T1-flows studied by \citet{dudley}, so that the dynamo action of the spin-up flow is not surprising. One can also attempt interpretations in terms of mean field magnetohydrodynamics with its two main ingredients, the $\alpha$ and $\Omega$ effects. There is undoubtedly differential rotation present in the spin-up flow, so that some $\Omega$ effect is plausible. In the limit of good scale separation and small magnetic Reynolds number on the scale of the smallest motion, the $\alpha$ effect is related to the helicity of flow. Figure \ref{heli} shows the spatial distribution of the helicity. It is anti-symmetric with respect to the equator, reverses sign after each collision, and is concentrated in the boundary layers. The helicity $h=\bm u\cdot(\bm\nabla\times\bm u)$ has three terms, $h=u_r(\bm\nabla\times\bm u)_r+u_\theta(\bm\nabla\times\bm u)_\theta+u_\phi(\bm\nabla\times\bm u)_\phi$. The first term turns out to be negligible compared to the other two terms, which are comparable to each other. The components of $\bm\nabla\times\bm u$ are of course largest in the boundary layers, so that helicity has to be large there, too. From the field pattern described above, it is not possible to discriminate between an $\alpha$-$\Omega$ dynamo and an $\alpha^2$ dynamo (with the $\alpha$ of opposite sign in the two hemispheres) whose field is distorted by the differential rotation without adding constructively to the induction. The observed field pattern is compatible with both possibilities. In all of our dynamos, the toroidal kinetic energy contributes to 90\% of the total kinetic energy and the toroidal magnetic energy to 60\% of the total magnetic energy, but this again cannot indicate which type of dynamo they belong to.

\begin{figure} 
\caption{Snapshot of longitude-averaged fluid flow and magnetic field in the meridional plane. From left to right the four panels are angular velocity, meridional circulation, the toroidal field and poloidal field. Solid lines denote positive (anti-clockwise), and dashed lines negative (clockwise).  (a) $Ek=2\times10^{-3}$, $Rm=3\times10^3$, $\tau=10$ and ${\rm mod}(IP(t/\tau),2)=0$. (b) $Ek=2\times10^{-3}$, $Rm=3\times10^3$, $\tau=10$ and ${\rm mod}(IP(t/\tau),2)=1$. (c) $Ek=5\times10^{-4}$, $Rm=10^4$, $\tau=10$ and ${\rm mod}(IP(t/\tau),2)=0$. (d) $Ek=5\times10^{-4}$, $Rm=10^4$, $\tau=10$ and ${\rm mod}(IP(t/\tau),2)=1$}\label{mer}
\centering 
\subfloat[$Ek=2\times10^{-3}$, $Rm=3\times10^3$, $\tau=10$ and ${\rm mod}(IP(t/\tau),2)=0$.]
{\includegraphics[scale=0.8]{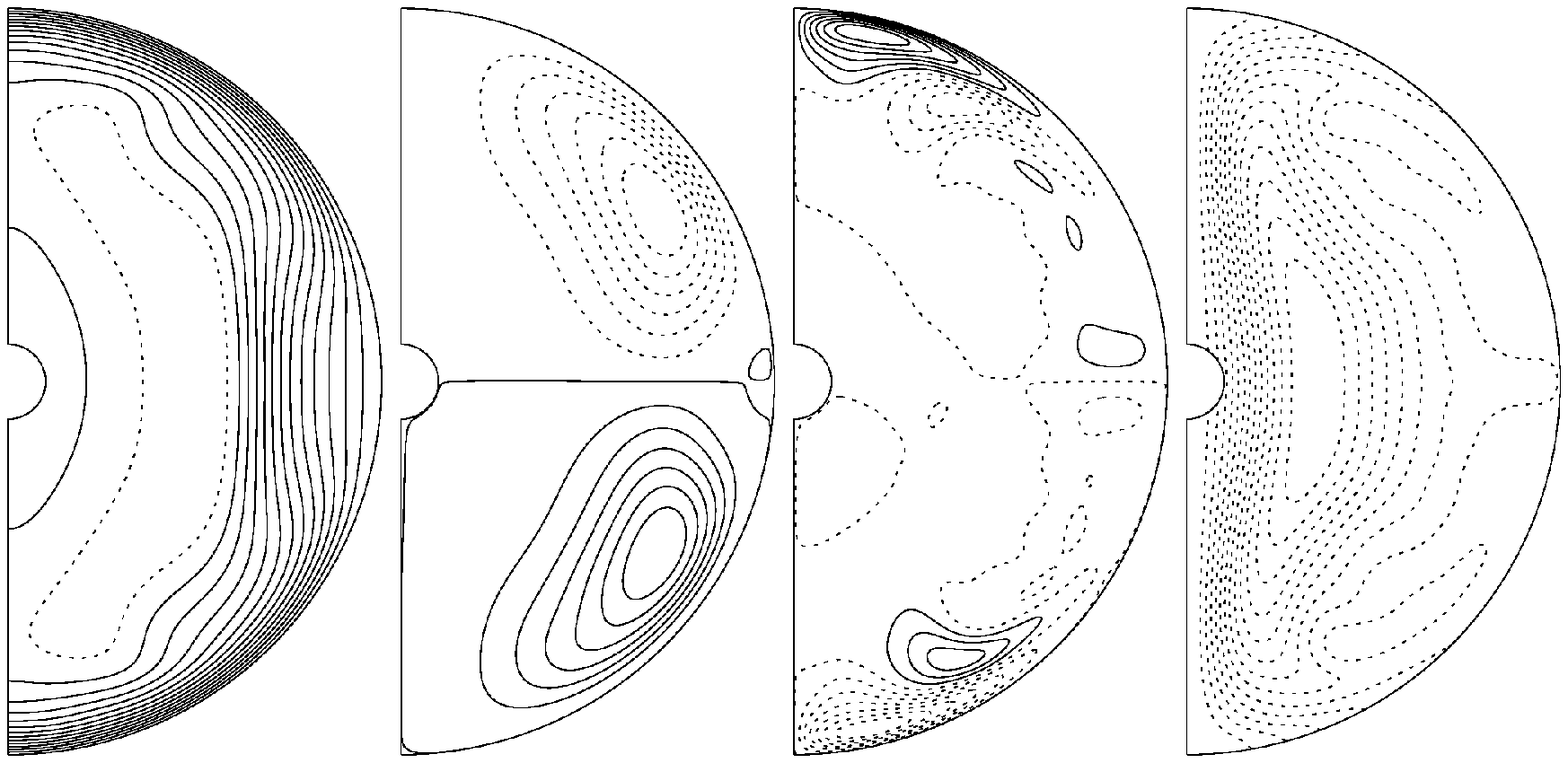}\label{mer1}}
\end{figure}

\begin{figure} 
\ContinuedFloat 
\centering 
\subfloat[$Ek=2\times10^{-3}$, $Rm=3\times10^3$, $\tau=10$ and ${\rm mod}(IP(t/\tau),2)=1$.]{\includegraphics[scale=0.8]{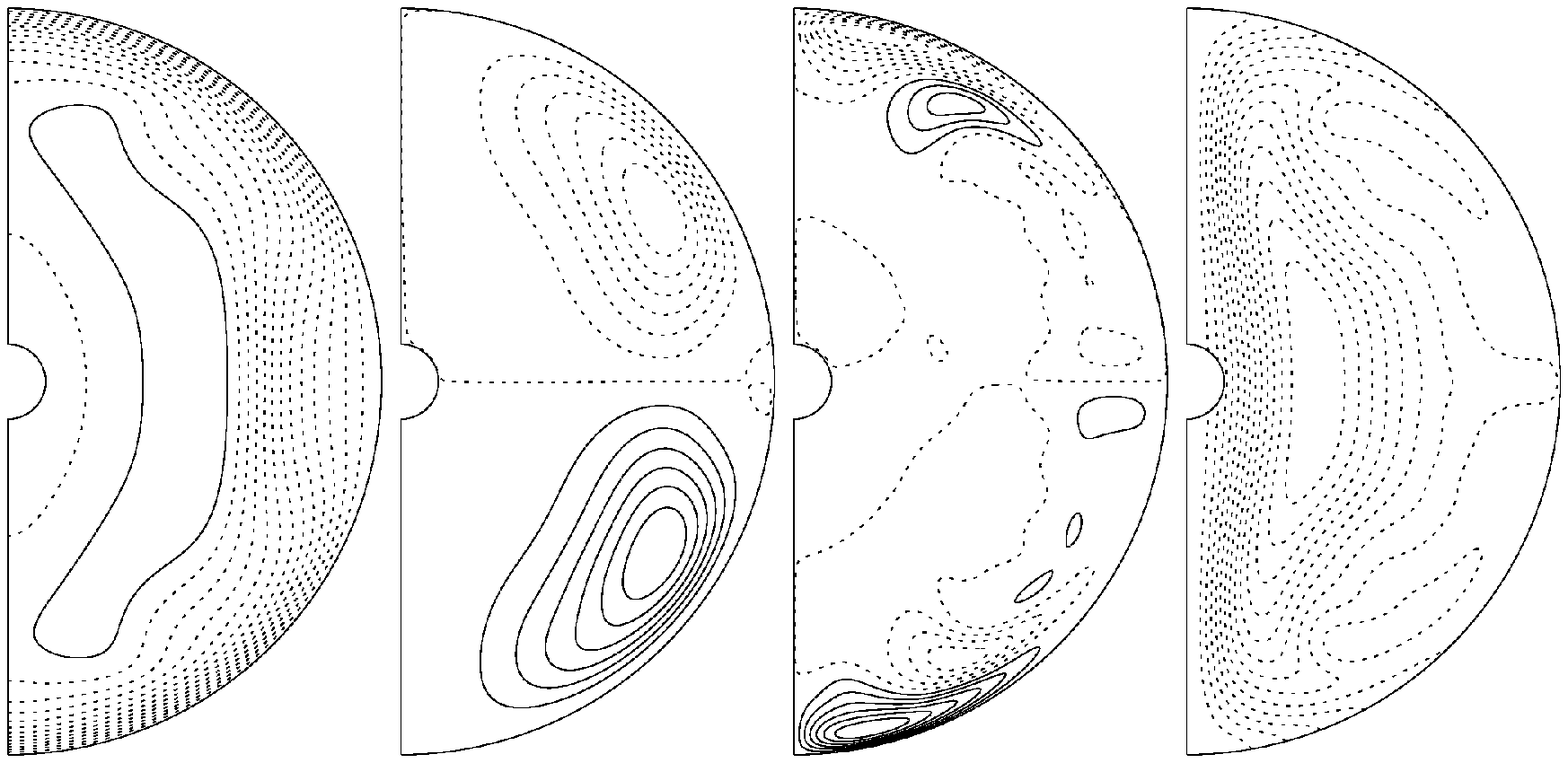}\label{mer2}}
\phantomcaption 
\end{figure}

\begin{figure} 
\ContinuedFloat 
\centering 
\subfloat[$Ek=5\times10^{-4}$, $Rm=10^4$, $\tau=10$ and ${\rm mod}(IP(t/\tau),2)=0$.]
{\includegraphics[scale=0.8]{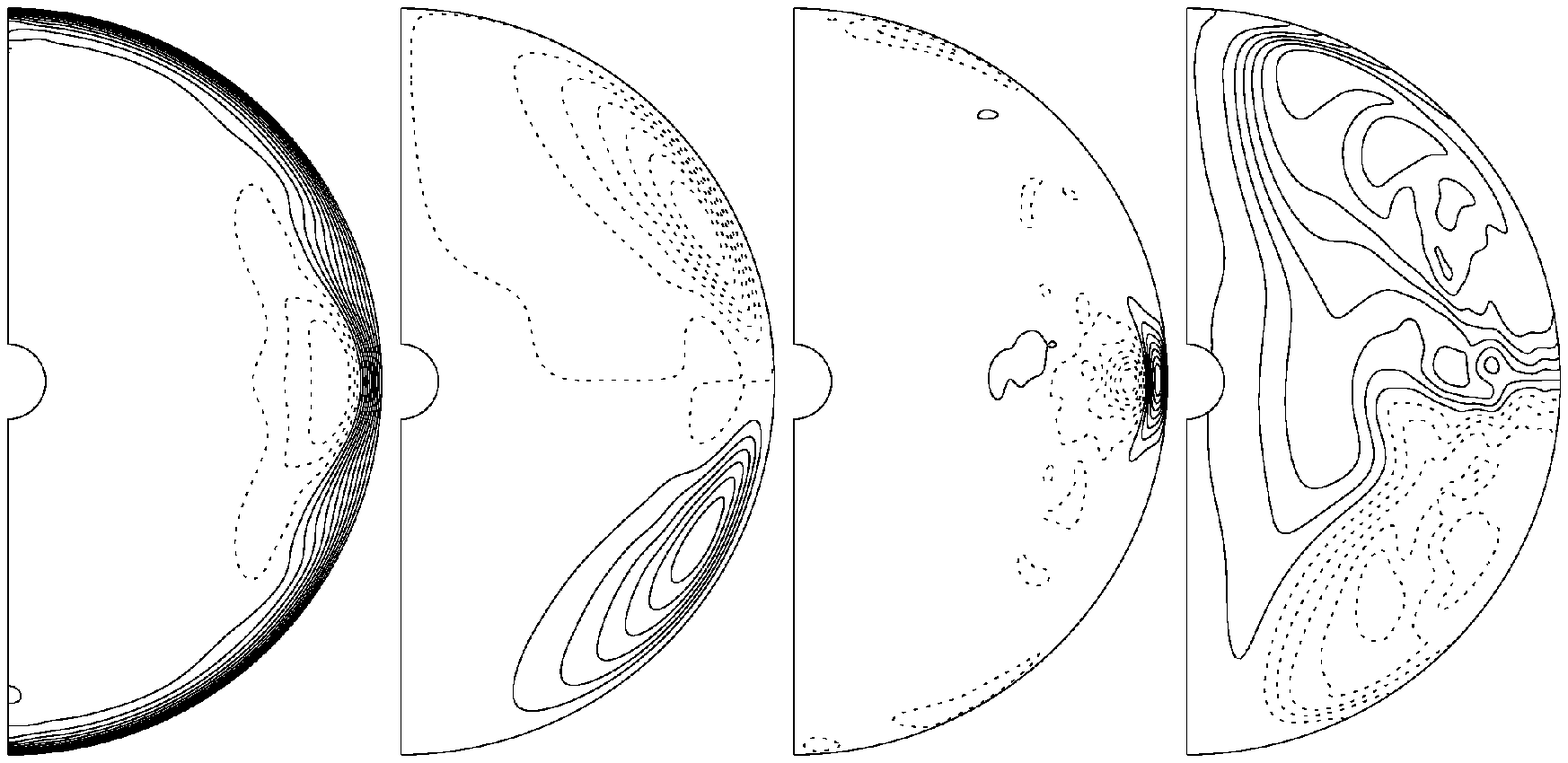}\label{mer3}}
\phantomcaption 
\end{figure}

\begin{figure} 
\ContinuedFloat 
\centering 
\subfloat[$Ek=5\times10^{-4}$, $Rm=10^4$, $\tau=10$ and ${\rm mod}(IP(t/\tau),2)=1$.]
{\includegraphics[scale=0.8]{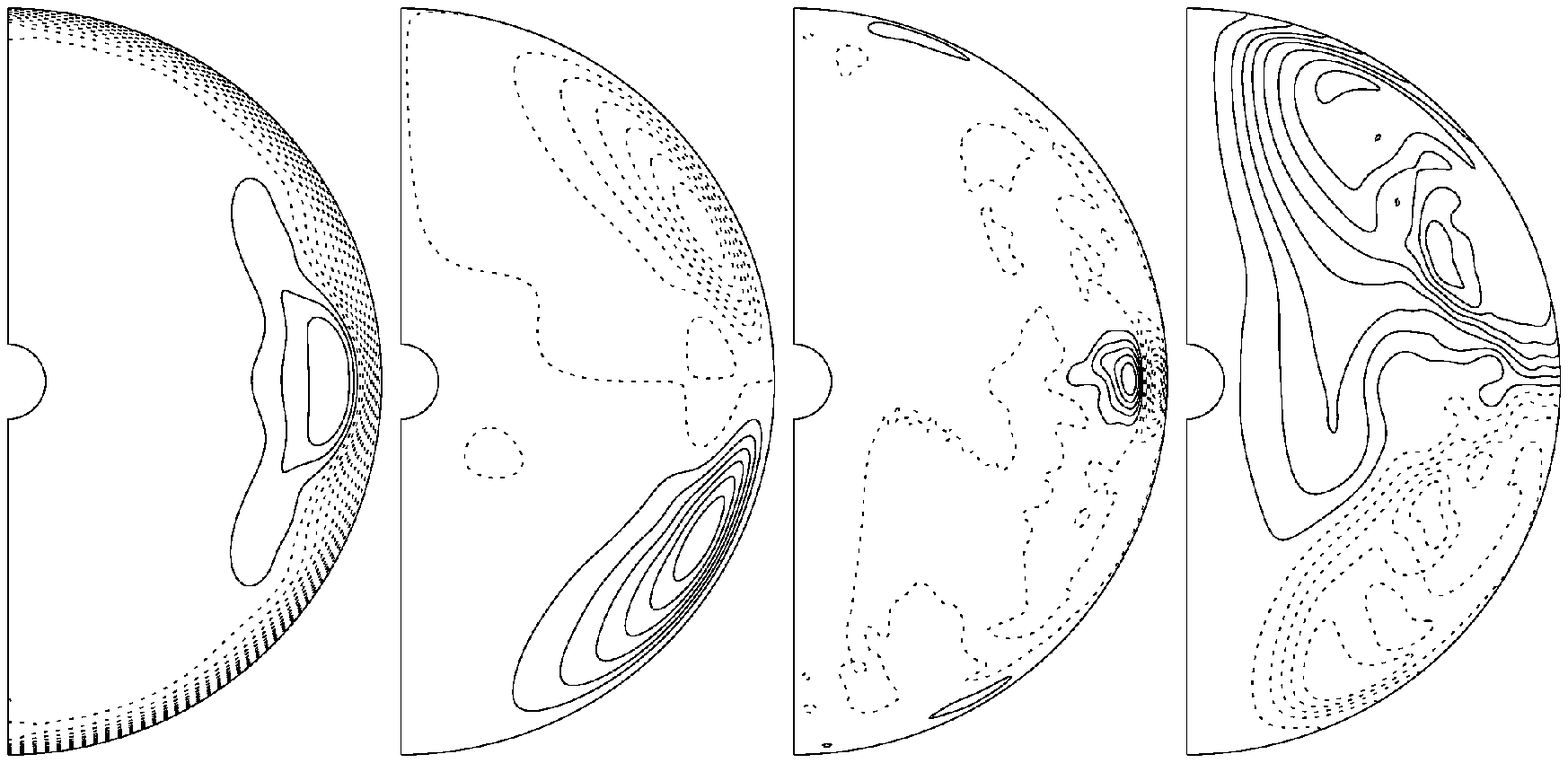}\label{mer4}}
\phantomcaption 
\end{figure}

\begin{figure} 
\centering 
\includegraphics[scale=0.8]{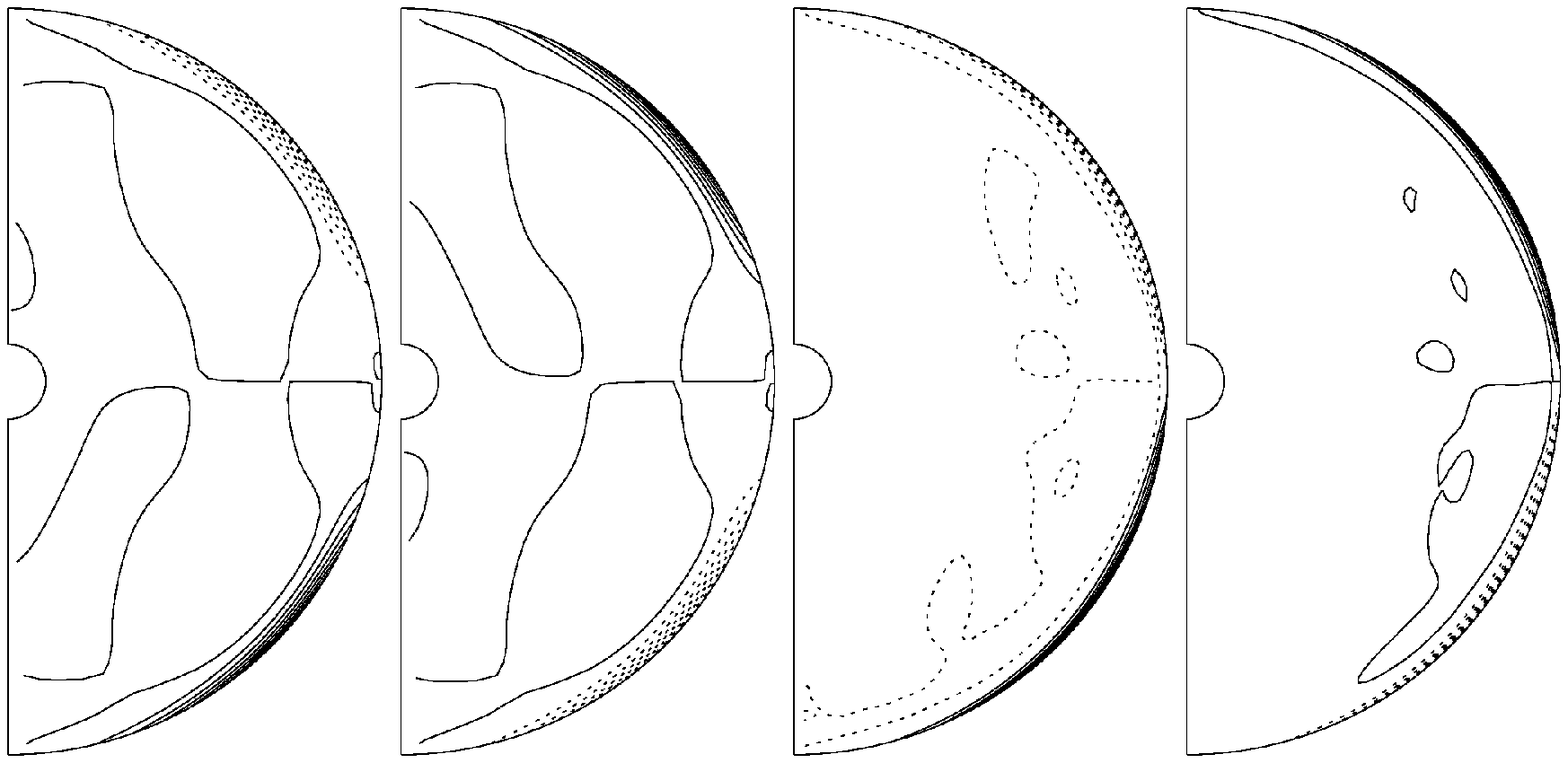}
\caption{Longitude-averaged helicity in the meridional plane. From left to right, the four panels correspond to the subfigures (a), (b), (c) and (d) in figure \ref{mer}.}\label{heli} 
\end{figure}

We now investigate the spectra of the velocity and magnetic fields. Figure \ref{spec} shows the power spectra of fluid velocity and magnetic field at the two Ekman numbers corresponding to figures \ref{mer} and \ref{heli}. The laminar spin-up flow realised at the larger of the two Ekman numbers is of course axisymmetric and symmetric with respect to the equator, as shown in subfigure \ref{spec1}. A kinematic dynamo with such a velocity field excites a magnetic field with a single azimuthal wavenumber $m$ and a well defined symmetry with respect to the equator. All calculations have been done including the Lorentz force, but in the situation close enough to the onset of dynamo instability, one still finds a magnetic field dominated by the $m=1$ mode, and the equatorial symmetry is reflected by the fact that the toroidal (poloidal) modes with even (odd) $l-m$ have much larger amplitudes than those with odd (even) $l-m$. At the lowest $Ek$ the equatorial symmetry in the velocity field is broken by the first instability as shown in subfigure \ref{spec2}. The magnetic field then also loses the symmetries it had in the laminar case. One expects a more complex and three dimensional flow to be a better dynamo than the axisymmetric laminar spin-up flow, and this instability may explain why the dynamo window at the lowest $Ek$ is much wider than the other three dynamo windows at the higher Ekman numbers in figuer \ref{rev}.

\begin{figure} 
\caption{Power spectra of fluid flow and magnetic field. The vertical axis has a logarithmic scale. $l$ is the colatitude wavenumber and $m$ the azimuthal wavenumber. From left to right, the top row shows the colatitude spectra of, respectively, toroidal flow $e$, poloidal flow $f$, toroidal magnetic field $g$ and poloidal magnetic field $h$ (as in equation \ref{t-p}). The bottom row shows the corresponding longitude spectra. (a) $Ek=2\times10^{-3}$, $Rm=3\times10^3$ and $\tau=10$ (corresponding to subfigures \ref{mer1} and \ref{mer2}). (b) $Ek=5\times10^{-4}$, $Rm=10^4$ and $\tau=10$ (corresponding to subfigures \ref{mer3} and \ref{mer4}).}\label{spec} 
\centering
\subfloat[$Ek=2\times10^{-3}$, $Rm=3\times10^3$ and $\tau=10$.]
{\includegraphics[scale=0.8]{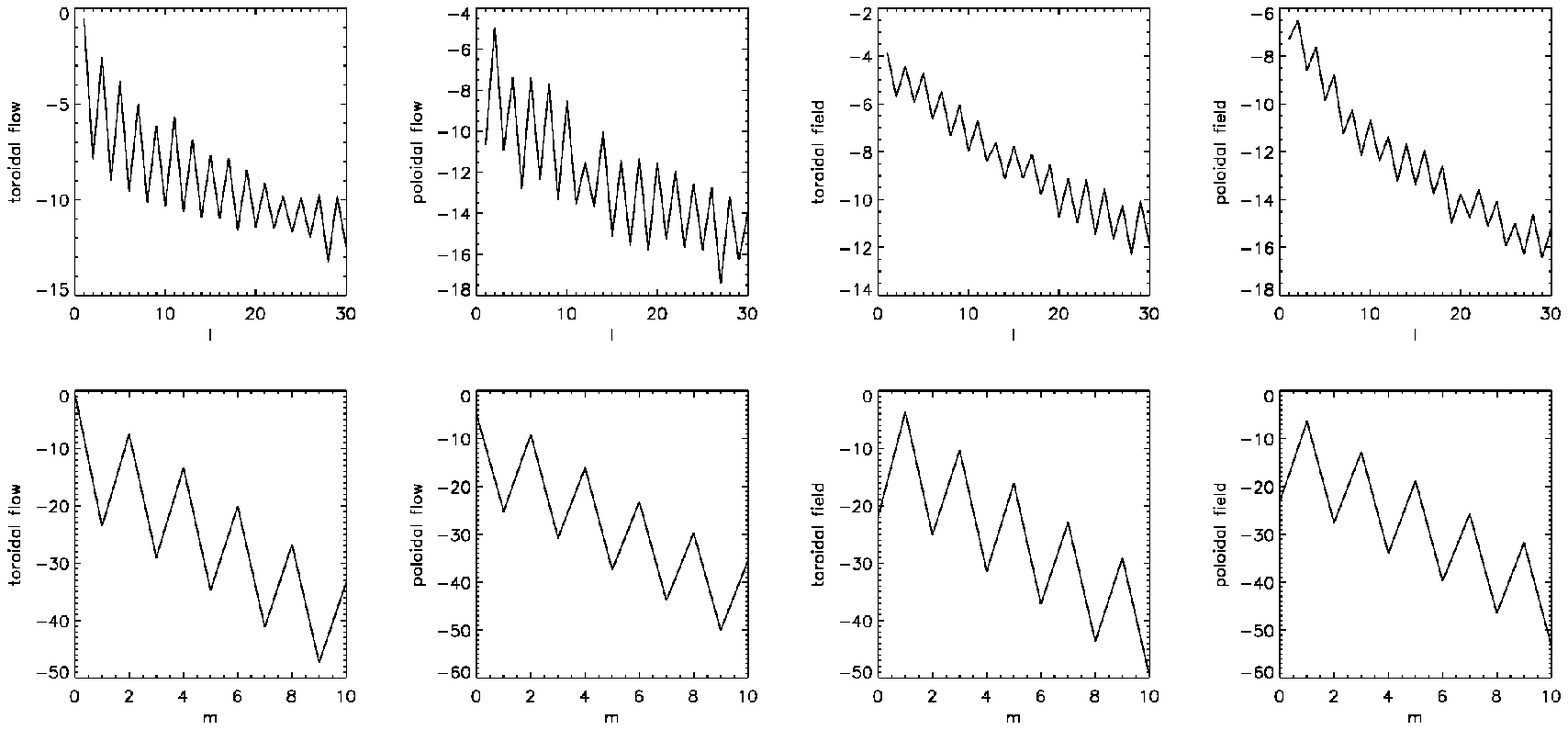}\label{spec1}} 
\end{figure}

\begin{figure} 
\ContinuedFloat 
\centering 
\subfloat[$Ek=5\times10^{-4}$, $Rm=10^4$ and $\tau=10$.]
{\includegraphics[scale=0.8]{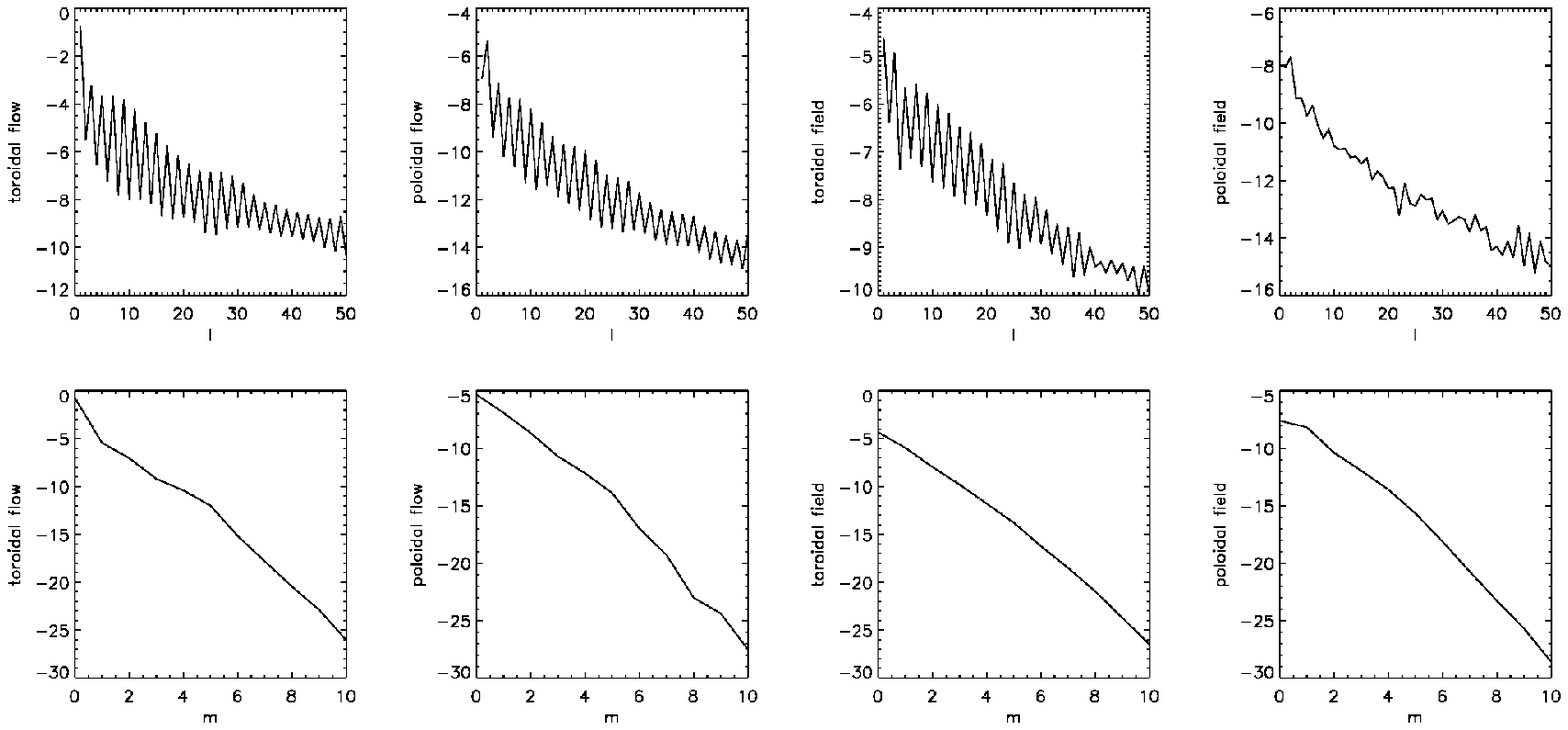}\label{spec2}}
\phantomcaption 
\end{figure}

\begin{figure} 
\centering 
\includegraphics[scale=0.7]{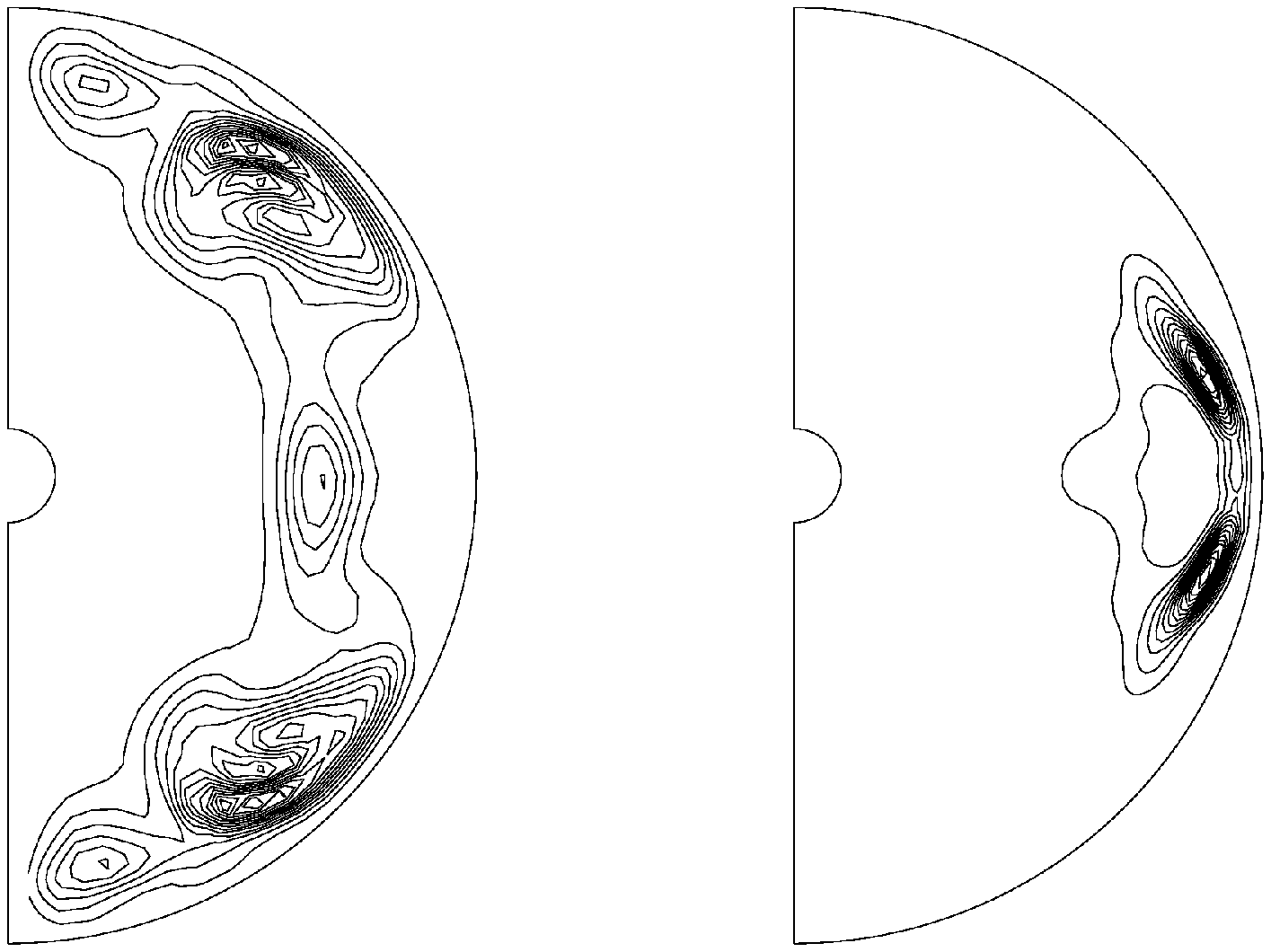}
\caption{Longtude-averaged kinetic energy of non-axisymmetric instabilites $\left(\left<\bm u\cdot\bm u\right>-\left<\bm u\right>\cdot\left<\bm u\right>\right)/2$ in the meridional plane. The left panel corresponds to subfigures \ref{mer1} and \ref{mer2} and the right panel to subfigures \ref{mer3} and \ref{mer4}.}\label{ekin} 
\end{figure}

Figure \ref{ins} shows time traces of different contributions to the total kinetic energy at two different Ekman numbers at $\tau=20$. At the higher $Ek$ (subfigure \ref{ins2}) the flow is not a dynamo and the field decays to zero whereas at the lowest $Ek$ (subfigure \ref{ins4}) it is a dynamo. At the higher Ekman number, the $m=1$ component of the flow is negligible and at the level of round-off errors. The same is true for the axisymmetric component with the equatorial symmetry opposite to the symmetry imposed by the boundary condition. At the lowest Ekman number, energy appears in all modes irrespective of azimuthal wavenumber and equatorial symmetry. The flow is axisymmetric at the three higher Ekman numbers in figure \ref{rev} where no dynamo action occurs at $\tau=20$, whereas the flow at the lowest $Ek$ is unstable and dynamos are observed up to $\tau=70$. Therefore, it is likely that the non-axisymmetric hydrodynamic instability is responsible for the wider dynamo window at the lowest $Ek$. We can also investigate the distribution of non-axisymmetric energy. Figure \ref{ekin} shows the distribution of kinetic energy of non-axisymmetric instabilities $\left(\left<\bm u\cdot\bm u\right>-\left<\bm u\right>\cdot\left<\bm u\right>\right)/2$ (where the brackets denote the azimuthal average) in the meridional plane. The location of instability corresponds to the location of the largest shear, as can be seen by comparing the left panel in figure \ref{ekin} to the first panel in subfigures \ref{mer1} and \ref{mer2} and the right panel in figure \ref{ekin} to the first panel in subfigures \ref{mer3} and \ref{mer4}, which are visualizations of the same fluid state.

\begin{figure} 
\caption{Kinetic energies on different modes as a function of time. (a) $Ek=2\times10^{-3}$ and $\tau=20$. (b) $Ek=5\times10^{-4}$ and $\tau=20$. Red for toroidal component $e$ of $(k=1,l=1,m=0)$, green for poloidal component $f$ of $(k=1,l=1,m=0)$, blue for toroidal component $e$ of $(k=1,l=1,m=1)$ and black for poloidal component $f$ of $(k=1,l=1,m=1)$.}\label{ins}
\centering 
\subfloat[$Ek=2\times10^{-3}$ and $\tau=20$.]{\includegraphics[scale=0.7]{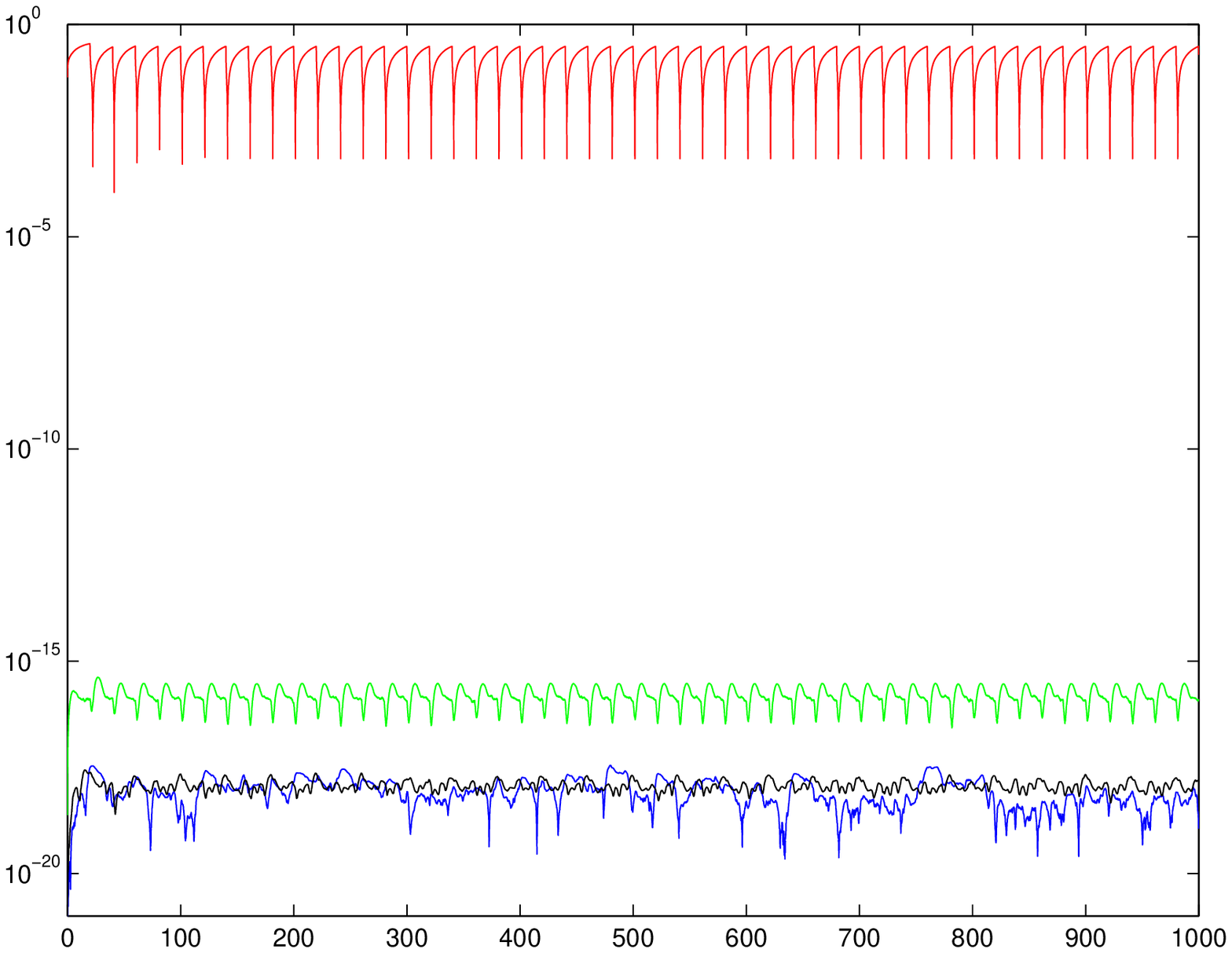}\label{ins2}} 
\end{figure}

\begin{figure} 
\ContinuedFloat 
\centering 
\subfloat[$Ek=5\times10^{-4}$ and $\tau=20$.]{\includegraphics[scale=0.7]{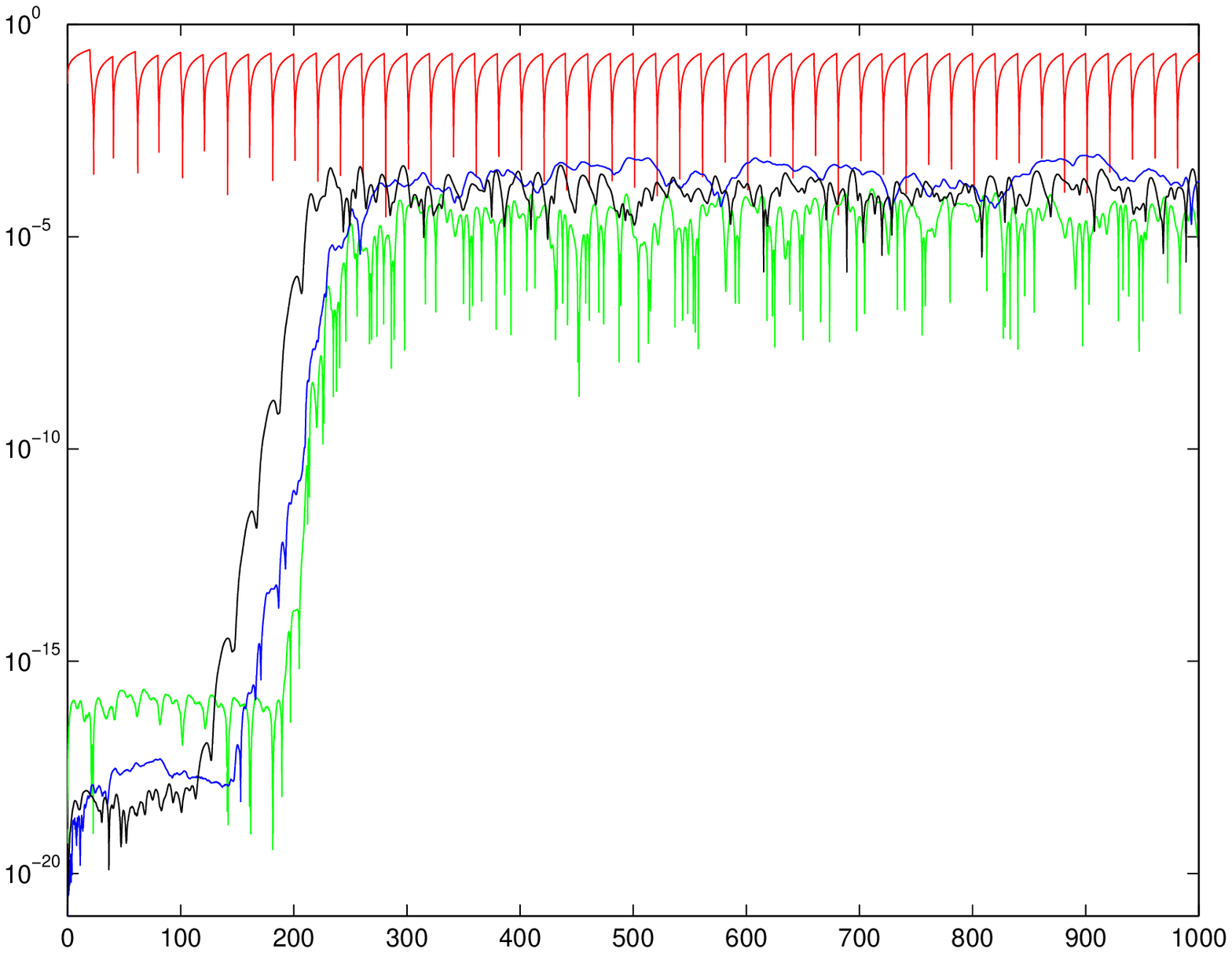}\label{ins4}}
\phantomcaption
\end{figure}

\begin{figure} 
\caption{Saturated magnetic energy as a function of $Rm$ at different $Ek$ and $\tau$. Red colour for $Ek=5\times10^{-3}$, green for $Ek=2\times10^{-3}$, blue for $Ek=1\times10^{-3}$ and black for $Ek=5\times10^{-4}$. Square for $\tau=5$, circle for $\tau=10$ and triangle for $\tau=20$.} 
\centering \includegraphics[scale=0.66]{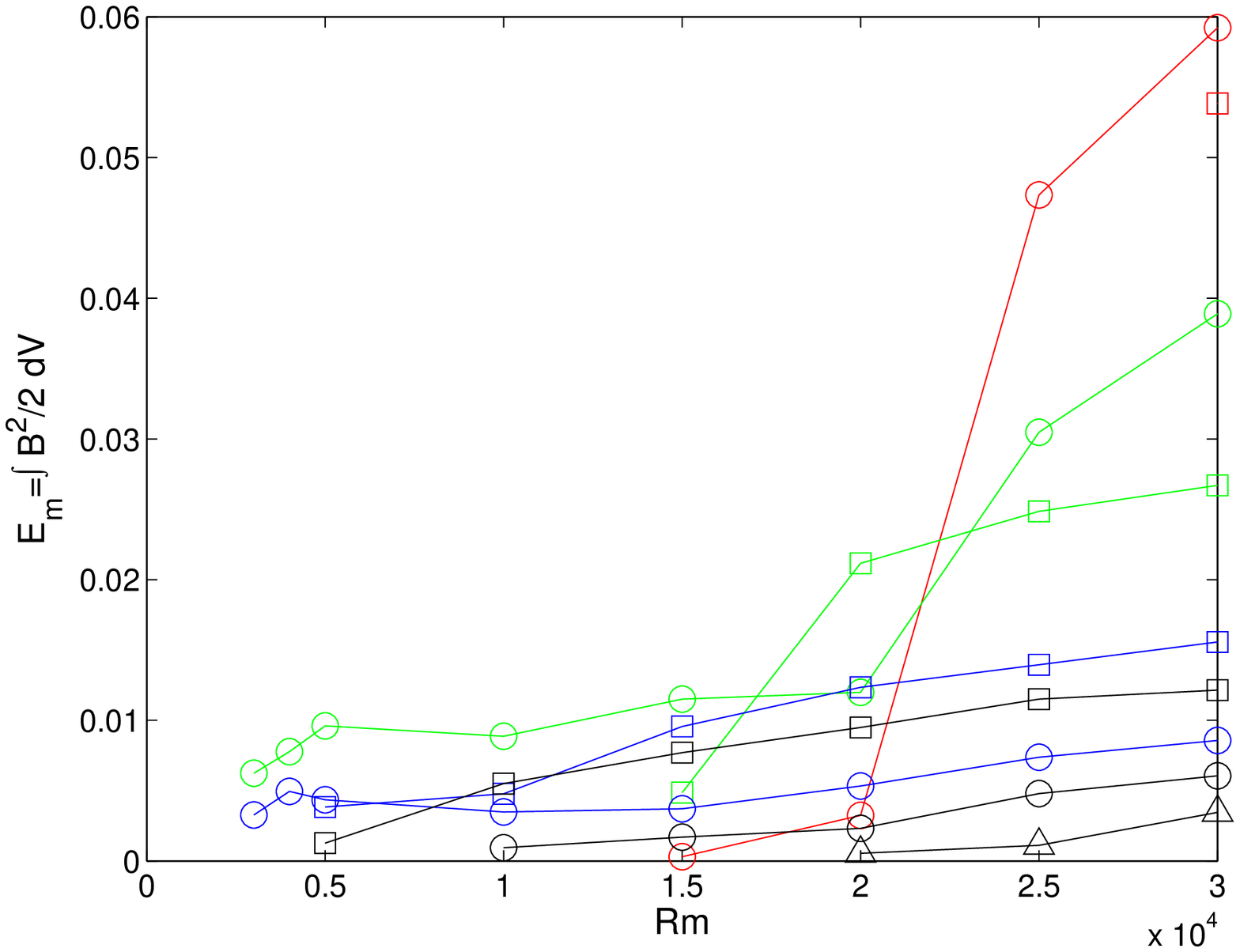}\label{em-rm}
\end{figure}

We then investigate the saturated magnetic energy integrated over the spherical shell. First of all, in our dynamo solutions, at the saturation level, the ratio of magnetic energy to kinetic energy is around $0.15$, which suggests that the spin-up dynamos in our parameter regime are weak field dynamos. Figure \ref{em-rm} shows the magnetic energy as a function of $Rm$ at different $Ek$ and $\tau$. It is reasonable that a higher $Rm$ corresponds to a higher magnetic energy. It is also reasonable that a lower $Ek$ corresponds to a lower magnetic energy because the spin-up flow at a low $Ek$ cannot fully develop and reach its maximum between two successive collisions. In the convection dynamo, $Ek$ determines the onset of convection and hence how far overcritical a certain $Ra$ is, it fixes the size of the convection rolls, and determines the Coriolis force in the magnetostrophic balance, which leads to various scaling laws of magnetic energy against $Ek$ \citep[e.g.][]{jones,christensen}. However, in the spin-up dynamo, $Ek$ is responsible for the energy input. A higher $Ek$ corresponds to a larger viscosity (at a constant rotation rate) and hence more power injection into the fluid. Therefore, in the spin-up dynamo the saturation of magnetic energy is largely determined by the time left until the next collision or until the spin-up process has finished, but not by a force balance as in the convection dynamo. Increasing $\tau$ at a constant $Rm$ allows the kinetic energy to grow further, but it also leaves more time for the magnetic field to decay. But it looks anomalous that at the highest $Ek=5\times10^{-3}$ there is an abrupt jump of magnetic energy between the low $Rm$ regime ($1.5-2\times10^4$) and the high $Rm$ regime ($2.5-3\times10^4$). There is also an abrupt jump in the non-axisymmetric kinetic energy at around $Rm=2\times10^4$. At the the highest $Ek$, the purely hydrodynamic flows with the absence of magnetic field are stable, but the hydromagnetic flows with the presence of magnetic field exhibit instability at high $Rm=25000-30000$, and this instability at $Ek=5\times10^{-3}$ leads to a higher magnetic energy. This type of instability that occurs when $Rm$ is increased might be caused by the magnetic field, i.e. the so-called magnetic instability \citep{acheson}.

To end this section we briefly investigate the intermittency of the magnetic field at the lowest $Ek=5\times10^{-4}$. Subfigure \ref{interm1} shows peaks in the time series of magnetic energy which typically last for about $1000$ units of time or $100$ collisions. The poloidal field keeps its direction during the intermittency. The dimensionless spin-up time scale at this lowest $Ek$ is $Ek^{-1/2}=45$ and the dimensionless magnetic diffusion time scale is $Rm=10000$. Both of them are far away from the intermittency time scale $1000$. Therefore, this intermittency is correlated neither to Ekman pumping nor to magnetic diffusion, but might be caused by some complicated nonlinear effect. Since the polarity does not reverse, the surface rocks of a planetesimal can be magnetised without cancellation. If one zooms into one of these peaks as shown in subfigure \ref{interm2}, one discovers smaller peaks separated by 10 units of time. These peaks are of course expected for $\tau=10$, but there is at present no explanation for the appearance of the long time scale $1000$. This intermittency is not observed at the higher Ekman numbers.

\begin{figure} \caption{Intermittency of magnetic energy. $Ek=5\times10^{-4}$,
$Rm=10^4$ and $\tau=10$.}\label{interm} \centering \subfloat[Magnetic energy as
a function of time.]{\includegraphics[scale=0.66]{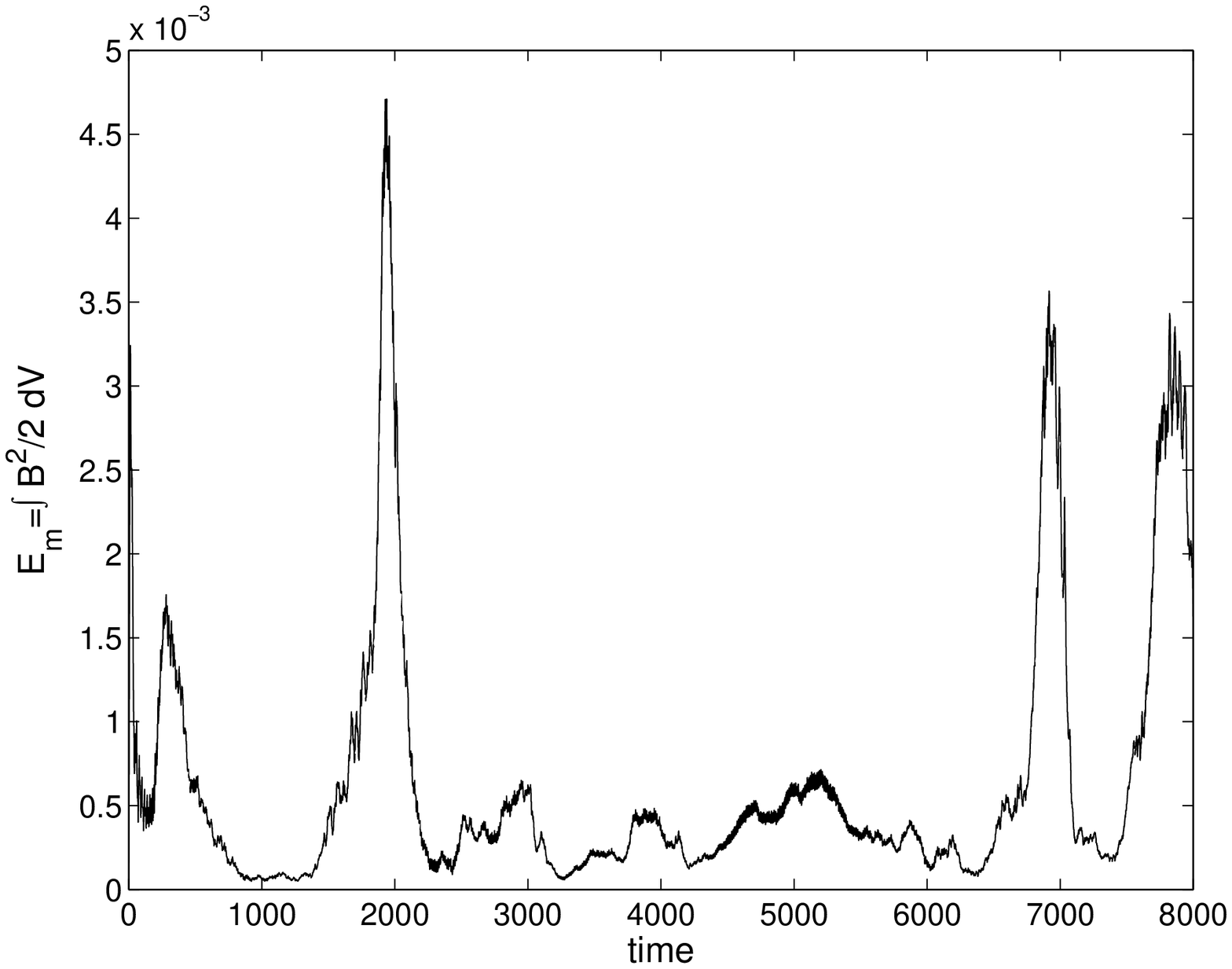}\label{interm1}}
\end{figure}

\begin{figure} \ContinuedFloat \centering \subfloat[Zoom of the peak at $t=1900
- 2000$.]{\includegraphics[scale=0.66]{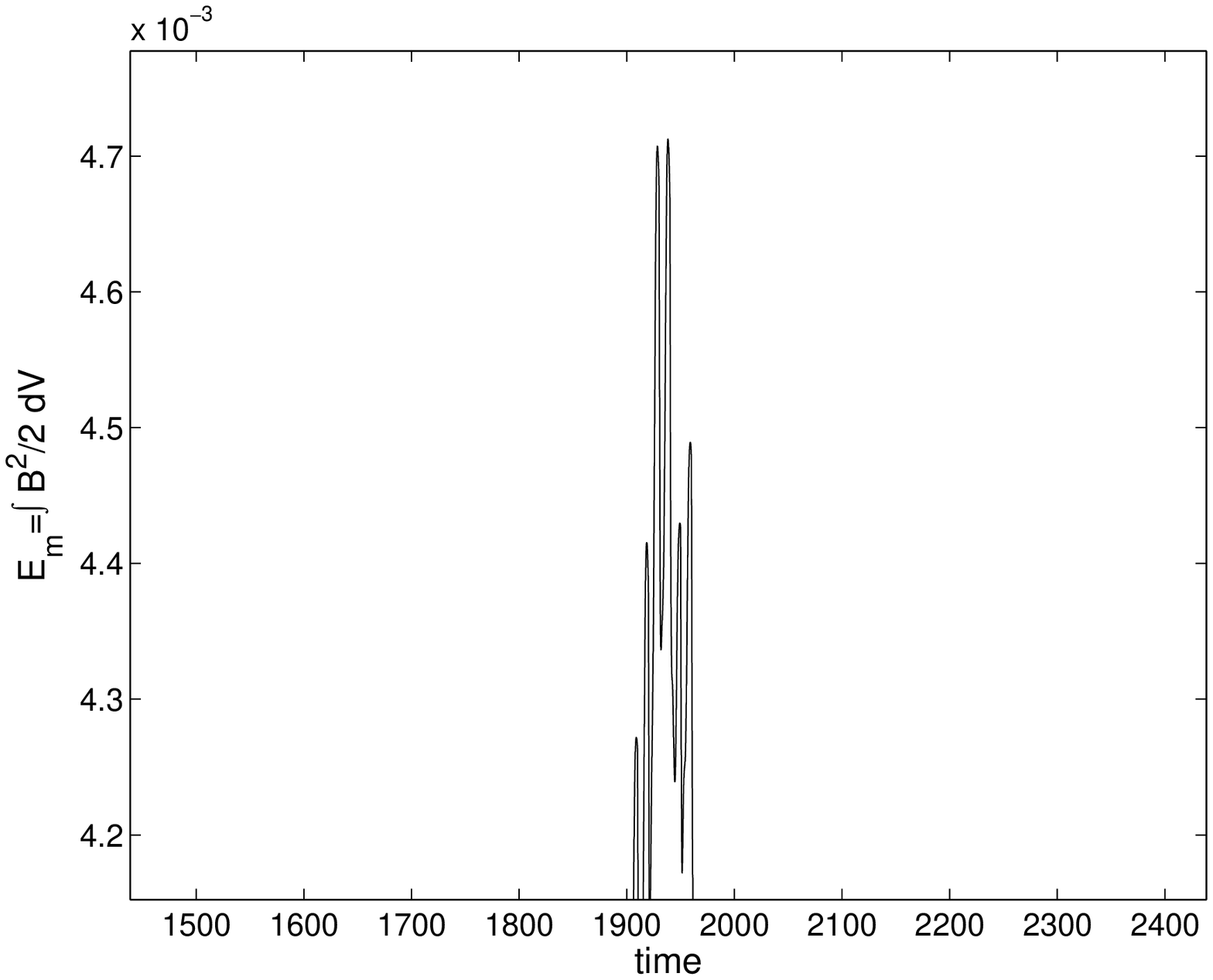}\label{interm2}}
\phantomcaption \end{figure}

\section{Discussion} 
We numerically calculate the nonlinear self-sustained dynamo in a spherical shell driven by the spin-up process due to the periodic reversals of the rotation of the outer boundary. We investigate different Reynolds numbers or Ekman numbers, magnetic Reynolds numbers and collision rates. It is found that there exists a dynamo window for collision rates and this window is wider at lower $Ek$. This spin-up flow is helical near the boundary. The toroidal field reverses after each collision whereas the poloidal field keeps its direction. A non-axisymmetric hydrodynamic instability at low $Ek$ helps dynamo action. The magnetic field is temporally intermittent at low $Ek$.

Although we use a simplified spin-up dynamo model to interpret the magnetic field of planetesimals and our numerical calculations are far away from the real parameter regime of the early solar system, it appears that collisions are a plausible energy source for dynamos in planetesimals.

{\small\section*{Acknowledgement}I am financially supported by the project SPP1488 of the program PlanetMag of Deutsche Forschungsgemeinschaft (DFG).}


\end{document}